\documentstyle[12pt]{article}

\def\d{\partial}
\def\dh{\mathop{\vphantom{\odot}\hbox{$\partial$}}}
\def\dl{\dh^\leftrightarrow}
\def\sqr#1#2{{\vcenter{\vbox{\hrule height.#2pt\hbox{\vrule width.#2pt
height#1pt \kern#1pt \vrule width.#2pt}\hrule height.#2pt}}}}
\def\w{\mathchoice\sqr45\sqr45\sqr{2.1}3\sqr{1.5}3\,}

\def\=d{\,{\buildrel\rm def\over =}\,}

\def\i3p{\p32\int d^3p}

\def\As{A\hbox to 1pt{\hss /}}
\def\np4{\int d^4p_1\cdots d^4p_{n-1}\, }

\def\nx4{\int d^4x_1\ldots d^4x_n\, }

\def\Dr{D^{\rm ret}}
\def\Da{D^{\rm av}}
\def\kon#1#2{\vbox{\halign{##&&##\cr
\lower4pt\hbox{$\scriptscriptstyle\vert$}\hrulefill &
\hrulefill\lower4pt\hbox{$\scriptscriptstyle\vert$}\cr $#1$&
$#2$\cr}}}

\def\konv#1#2#3{\hbox{\vrule height12pt depth-1pt}
\vbox{\hrule height12pt width#1cm depth-11.6pt}
\hbox{\vrule height6.5pt depth-0.5pt}
\vbox{\hrule height11pt width#2cm depth-10.6pt\kern5pt
      \hrule height6.5pt width#2cm depth-6.1pt}
\hbox{\vrule height12pt depth-1pt}
\vbox{\hrule height6.5pt width#3cm depth-6.1pt}
\hbox{\vrule height6.5pt depth-0.5pt}}
\def\konu#1#2#3{\hbox{\vrule height12pt depth-1pt}
\vbox{\hrule height1pt width#1cm depth-0.6pt}
\hbox{\vrule height12pt depth-6.5pt}
\vbox{\hrule height6pt width#2cm depth-5.6pt\kern5pt
      \hrule height1pt width#2cm depth-0.6pt}
\hbox{\vrule height12pt depth-6.5pt}
\vbox{\hrule height1pt width#3cm depth-0.6pt}
\hbox{\vrule height12pt depth-1pt}}

\def\konw#1#2#3{\hbox{\vrule height12pt depth-1pt}
\vbox{\hrule height12pt width#1cm depth-11.6pt}
\hbox{\vrule height6.5pt depth-0.5pt}
\vbox{\hrule height12pt width#2cm depth-11.6pt \kern5pt
      \hrule height6.5pt width#2cm depth-6.1pt}
\hbox{\vrule height6.5pt depth-0.5pt}
\vbox{\hrule height12pt width#3cm depth-11.6pt}
\hbox{\vrule height12pt depth-1pt}}

\def\eh{{\scriptstyle{1\over 2}}}

\def\i{{\rm int}}

\def\m3{{\mu_1\mu_2\mu_3}}

\def\p{{(+)}}

\begin{document}

\thispagestyle{empty}
{\it Z\"urich-University-Preprint ZU-TH-20/95}\\
{\bf}\\

\vbox to 2,5cm{ }
\centerline{\Large\it Nonabelian Gauge Symmetry }
\vskip 0.3cm
\centerline{\Large\it in the Causal Epstein-Glaser Approach }
\vskip1.5cm
\centerline{\large\it Tobias Hurth$^*$}
\vskip 0.5cm
\centerline{\large\it
Institute for Theoretical Physics, University of Z\"urich}
\centerline{\large\it Winterthurerstr. 190, CH-8057 Z\"urich, Switzerland}
\vskip 1,0cm
{\bf Abstract.} - We present some generalizations of a recently proposed
alternative approach to nonabelian gauge theories based on the causal
Epstein-Glaser method in perturbative quantum field theory.
\\
Nonabelian gauge invariance is defined by a simple commutator relation in every
order of perturbation theory separately, using only the linear (abelian)
BRS-transformations of the asymptotic fields. This condition is sufficient for
the unitarity of the S-matrix in the physical subspace.
\\
We derive the most general specific coupling compatible with the conditions of
nonabelian gauge invariance and normalizability.
\\
We explicitly show that the quadrilinear terms, the four-gluon-coupling and the
four-ghost-coupling, are generated by our linear condition of nonabelian gauge
invariance.
Moreover, we work out the required generalizations for linear gauges.
\\

chromodynamics
\vskip 0.5cm
$^*)$ Emailaddress: hurth@physik.unizh.ch\\
$^*)$ Supported by the Swiss National Science Foundation\\
\vskip 0.3cm

\vfill\eject

\newpage
{\large\bf 1. Introduction}
\vskip1cm

The causal Epstein-Glaser formalism [1] is an alternative approach to
(perturbative) quantum field theory. The $S$-matrix is constructed in the
well-defined Fock space of free asymptotic fields in the form of a formal power
series

$$S(g)=1+\sum_{n=1}^{\infty}{1\over n!}\int d^4x_1...
d^4x_n\,T_n(x_1,...,x_n)g(x_1)...g(x_n),\eqno(1.1)$$
where $g(x)$ is a tempered test function which switches the interaction.
Only well-defined free field operators occur in the whole construction.
Thus, one does not need the Haag-Ruelle (LSZ-) formalism. However, interacting
field operators can be perturbatively constructed in an additional step as
certain functional derivatives of the $S$-matrix [2].\\
The central objects are the n-point distributions $T_n$. They may be viewed as
mathematically well-defined time-ordered products.
\\
The defining equations of the theory in the causal formalism are the
fundamental (anti-) commutation relations of the free field operators, their
dynamical equations and the specific coupling of the theory $T_{n=1}$. The
$n$-point distributions $T_n$ in (1.1) are then constructed inductively from
the given first order $T_{n=1}$. In fact, Epstein and Glaser [1] present an
explicit inductive construction of the general perturbation series in the sense
of (1.1) which is compatible with causality and Poincare invariance. It is the
physical condition of causality which makes a direct construction of the
renormalized (finite) perturbation series possible - without any intermediate
modification of the theory and without introduction of any new mathematical
concept.
The whole perturbative $S$-matrix is already determined by the conditions of
causality, Poincare invariance and the specific coupling of the theory
$T_{n=1}$ except for a number of free constants which have to be fixed by
further physical conditions. \\
In the causal formalism the purely technical details which are essential for
explicit calculations are separated from the simple physical structure of the
theory:\\
The well-known ultraviolet (UV-)problem is reduced to a conceptionally simple
and mathematically well-defined problem, namely the splitting of an
operator-valued distribution with causal support into a distribution with
retarded and a distribution with advanced support. The physical infrared
(IR-)problem is naturally separated by the adiabatic switching of the
$S$-matrix $S(g)$ with a tempered test function $g$ and does only arise if one
considers the adiabatic limit
$g \rightarrow 1$ in (1.1).
\\
As a consequence, general properties of the perturbative quantum field theory
like normalizability, gauge invariance, unitarity or the infrared behaviour of
the theory can be separately analyzed and inductively proven.
\\
With this background the causal analysis of the abelian gauge theory in the
four dimensional Minkowski space was worked out ([3,4,5]). The causal
Epstein-Glaser construction of a nonabelian gauge theory in four(3+1)
dimensional space-time was done for the Yang-Mills theory with Faddeev-Popov
coupling and with fermionic matter fields in the Feynman gauge [6,7]:\\
 A new definition of nonabelian gauge invariance is given as a simple
commutator condition in every order of perturbation theory separately.
For this purpose one only needs the concept of the linear (abelian)
BRS-transformations of the free asymptotic field operators.
The condition implies  unitarity of the $S$-matrix in the physical subspace,
i.e.  decoupling of the unphysical degrees of freedom. Moreover, one can derive
the well-known Slavnov-Taylor identities.\\
It is remarkable that the  content of nonabelian gauge invariance in
perturbation theory can be completely discussed in the well-defined Fock space
of  free asymptotic fields.\\
Furthermore, one can imagine, that the analysis of nonabelian gauge invariance
in perturbation theory is simplified considering the facts that in the usual
Lagrange formalism the BRS-transformations of the interacting fields [8] - the
central objects in this approach - relate basic fields with composite fields
and  that the latter transformations connect different orders of perturbation
theory.
\\
The Epstein-Glaser method is also well-suited to analyze genuine massive
nonabelian gauge theories [9]. The causal formalism allows for a comprehensive
and simplified analysis of such models  [10].\\
\\
The paper is organized as follows: In section 2 we present the
Epstein-Glaser construction of the Faddeev-Popov theory with fermionic matter
fields in four(3+1) dimensional space-time.
In section 3 we generalize these results to the most general
gauge invariant theory. In section 4 we discuss general $\xi-$gauges. In
Appendix A we prove Lemma 3.1 . In Appendix B, we give a brief introduction
into the causal Epstein-Glaser formalism.\\

\newpage

{\large\bf 2. A Linear Condition of Nonabelian Gauge Invariance}

\vskip1,0cm
In this section we present the causal construction of the (massless) Yang-Mills
theory with Faddeev-Popov coupling and with fermionic matter fields in four
(3+1)-dimensional space-time. We state the main theorems. For proofs we refer
to [6,7].\\
The corresponding specific coupling in the Feynman gauge is
$$T_1=igf_{abc}({1\over 2}:A_{\mu a}A_{\nu b}F^{\nu \mu}_c:
-:A_{\mu a}u_b\d^\mu\tilde u_c:)$$
$$+i\frac{g}{2}:\bar \psi_\alpha (\lambda_a)_{\alpha \beta}\gamma^\mu
\psi_\beta:A_{\mu a}.\eqno(2.1)$$
All the field operators herein are well-defined free fields and these are the
only quantities appearing in the whole theory. Double dots denote their normal
ordering.\\
The second term, the gluon-ghost-coupling, is naturally introduced by our
linear condition of nonabelian gauge invariance (see equations (2.14),(2.17)
below).\\
The specific coupling $T_{n=1}(x)$ of the theory does contain no quadrilinear
term proportional to $g^2$. Such terms are  automatically generated in second
order by our gauge invariance condition (see equation (2.17) below) as we
explicitly show in section 3. \\
$A_{\mu a}(x)$ are the (free) gauge potentials satisfying the commutation
relations (Feynman gauge)
$$[A_a^{(-)\mu}(x),A_b^{(+)\nu}(y)]=i\delta _{ab}
g^{\mu\nu}D_0^{(+)}(x-y),\eqno(2.3)$$
where $A^{(\pm)}$ are the emission and absorption parts of $A$ and
$D^{(\pm)}_0$ are the zero-mass Pauli-Jordan distributions. $u_a(x)$ and
$\tilde u_a(x)$ are the free massless fermionic ghost fields fulfilling the
anti-commutation relations
$$\{u_a^{(\pm)}(x),\tilde u_b^{(\mp)}(y)\}= -i\delta
_{ab}D_0^{(\mp)}(x-y).\eqno(2.4)$$
The gauge fields are minimally coupled to spinor fields
$\psi_\alpha,\bar{\psi}_\beta$. The latter satisfy the anti-commutation
relation
$$ \{\psi_\alpha^{(-)}(x), \bar{\psi}_\beta^{(+)}(y)\}=\delta_{\alpha \beta}
\frac{1}{i} S^{(+)} (x-y) \eqno(2.5)$$
where $S^{(+)}=(i\gamma_\mu \d^\mu + m) = D_m^{+}$. $f_{abc}$ denotes the usual
antisymmetric structure
constants of the gauge group, say $SU(N)$; $(-i/2) \lambda_a$ are the
generators of the fundamental representation of the Lie algebra of the gauge
group. The time-dependence of $A,u$ and $\tilde u, \psi$ and $\bar{\psi}$ is
given by the
wave equations
$$\w A_a^\mu(x)=0,\quad \w u_a(x)=0,\quad \w \tilde
u_a(x)=0, \eqno(2.6)$$
respectively by the free Dirac equation
$$i\gamma_\mu \d^\mu \psi_\alpha(x)=M_{\alpha \beta} \psi_\beta(x)
\eqno(2.7)$$
with a real and diagonal mass matrix $M_{\alpha \beta}$ which satisfies
$\lambda_a M = M \lambda_a $. For a simple gauge group like $SU(N)$ Schur`s
lemma implies that $M_{\alpha \beta}$ is a multiple of the unit matrix (2.5).
In the following we omit colour indices for the spinor fields in order to
simplify the notation. We define
$$F_a^{\mu\nu}\=d \d^\mu A_a^\nu -\d^\nu A_a^\mu, \quad j_a^\mu = : \bar{\psi}
\gamma^\mu \lambda_a \psi:. \eqno(2.8)$$
According to (2.7), $j_a^\mu$ is a conserved current: $ \quad \d_\mu j_a^\mu =
0.$
 \\
\\
Now one considers the linear (abelian!) BRS transformations [8,11] of the free
asymptotic field operators. The generator of the abelian operator
transformations is the charge
$$Q\=d \int d^3x\,(\d_{\nu}A_a^{\nu}{\dl}_0u_a), \quad Q^2=0,
\eqno(2.9)$$
with the (anti-)commutation relations
$$[Q,A_\mu^a]_-=i\d_\mu u_a,\quad \{ Q,\tilde{u}_a \}_+ = -i\d_\nu
A_a^\nu,\quad \{ Q,u_a \}_+ =0,$$
$$[Q,\psi]_-=0,\quad [Q,\bar{\psi}]_-=0,\quad [Q,F_{\mu \nu}^a]_-=0.
\eqno(2.10)$$
In addition to the charge $Q$, one defines the ghost charge
$$Q_c:=i \int d^3x :(\tilde u \stackrel{\leftrightarrow}{\d}_0 u):
\eqno(2.11)$$
In the algebra, generated by the fundamental field operators, one introduces a
gradation by the ghost number $G (\hat A)$ which is given on the homogenous
elements by
$$[Q_c, \hat A]= - G(\hat A) \cdot \hat A. \eqno(2.12)$$
One can define an anti-derivation $d_Q$ in the graded algebra by
$$d_Q \hat A:= Q \hat A-(e^{i\pi Q_c} \hat A e^{-i\pi Q_c}) Q\eqno(2.13)$$
The anti-derivation $d_Q$ is obviously homogenous of degree (-1) and satisfies
$d_Q^2 = 0$.\\
\\
Nonabelian gauge invariance in the causal approach means that the commutator of
the specific coupling (2.1) with the charge $Q$ is a divergence (in the sense
of vector analysis):
$$[Q,T_{n=1}]= i \d_\nu [igf_{abc} (:A_\nu^a u_b F_c^{\nu \mu} :- \eh : u_a u_b
\d^\nu \tilde{u}_c:)+igj_a^\nu u_a] \=d i\d_\nu T_{1}^\nu \eqno(2.14)$$
The second term in (2.1) (the gluon-ghost-coupling) is essential for that
 $d_Q T_{n=1}$ can be written as a divergence.
Note this  different compensation
of terms in the invariance equation (2.14) compared with the invariance of the
Yang-Mills Lagrangean under the full BRS-transformations of the interacting
fields in the conventional formalism where the gauge boson part ($TrFF$) alone
is BRS invariant.\\
The representation of $[Q, T_{n=1}]$ as a divergence is not unique in general.
The most general (so-called) $Q$-vertex $\tilde T_{1}^\nu$ with the same mass
dimension and ghost number as $T_{1}^\nu$ in (2.14) is the following:
$$[Q,T_1]=i\d_\nu [T_{1}^\nu + \gamma B_{1}^\nu] \=d i \d_\nu
\tilde{T}_{1}^\nu$$
$$\mbox{with} \qquad B_{1}^\nu=igf_{abc} \d_\mu (: u_a A_b^\mu A_c^\nu:), \quad
\d_\nu B_{1}^\nu =0, \quad \gamma \in \mbox{ {\bf C} } \quad \mbox{free.}
\eqno(2.15)$$
The choice of $\gamma$ has just practical reasons and has no physical
consequences.\\
In addition, $T_{n=1}$ in (2.1) is also anti-gauge invariant in the
sense that
$$ [ \bar Q , T_{n=1}] \quad ( \mbox{where} \quad \bar Q := \int
d^3x\,(\d_{\nu}A_a^{\nu}{\dl}_0 \tilde u_a) \quad \mbox{with} \quad \bar Q^2=0
) \quad $$
is also a divergence:\\
$$[\bar Q, T_{n=1}]_-= i\d_\nu[igf_{abc}(:\tilde u_a A_\kappa^b F_c^{\kappa
\nu}:-:A_\nu^a \d_\kappa A_b^\kappa \tilde u_c: +: \tilde u_a \d_\nu u_b \tilde
u_c:$$
$$-\eh \d_\nu(:\tilde u_a u_b \tilde u_c:) + igj_a^\nu \tilde
u_a]\quad+\quad\beta\quad i\d_\nu [igf_{abc} \d_\mu(:A_a^\nu \tilde u_b A_c^\mu
:)]$$
$$\=d i\d_\nu [\bar T_{1}^\nu + \beta \bar B_{1}^\nu]. \eqno(2.16)$$
Having defined nonabelian gauge invariance in the first order of perturbation
theory by the relation (2.14), the condition of nonabelian operator gauge
invariance in the causal approach is similarly
expressed in every order of perturbation theory separately by a simple
commutator relation of the n-point distributions $T_n$ with the charge $Q$, the
generator of the free operator gauge transformations:\\
\\
{\bf Theorem 2.1 (Nonabelian Gauge Invariance) } In the Faddeev-Popov theory in
(3+1)dimensional space-time with the defining equations (2.1-2.7),
the following linear condition holds in every order of perturbation theory:
$$[Q,T_n(x_1,...,x_n)]= d_Q  T_n(x_1,.....,x_n)= i\sum_{l=1}^n\d_\mu^{x_l}T^
\mu_{n/l}(x_1,...,x_n),\eqno(2.17)$$
where $T_{n/l}^\nu (x_1, \ldots, x_n)$ are n-point distributions of an extended
theory whose first order S-matrix is equal to
$$S_1(g_0,g_1)\=d \int d^4x \,[T_1(x)g_0(x)+
T_{1}^\nu (x)g_{1\nu}(x)].\eqno(2.18)$$
$g_1=(g_{1\nu})_{\nu=0,1,2,3}\in ({\cal S}({\bf R}^{\it 4}))^4$ must be an
anti-commuting C-number field. The higher orders are determined by the usual
inductive Epstein Glaser construction up to local normalization terms.
The $T^\mu_{n/l}$ are the n-point
distributions of the extended theory with one $Q$-vertex at $x_l$, all other
$n-1$ vertices are ordinary Yang-Mills vertices (2.1).\\
\\
The simple linear operator condition (2.17) involving only well-defined
asymptotic field operators expresses the full content of the nonabelian gauge
structure of the quantized theory in perturbation theory, namely the
Slavnov-Taylor identities and the unitarity of the S-matrix in the physical
subspace, i.e. the decoupling of the unphysical degrees of freedom in the
theory . It can be proven by induction on the order n of perturbation theory
following the causal construction of $T_n$ and $T_{n/l}^\nu$ [6,7]: The linear
operator condition
is expressed by a set of identities between C-number distributions analogously
to the Slavnov-Taylor identities. The different types of these identities are
derived and proven by suitable normalization. All symmetries of the theory,
in particular the global $SU(N)$-symmetry and charge conjugation invariance,
are needed. But in order to express the operator gauge invariance condition
(2.17) in a set of identities between C-number distributions,
one has to work out the explicit form of the divergence in the operator gauge
invariance condition. Moreover, one has to distinguish the operator and its
derivative , which implies the relative largeness of the set of identities to
be proven separately.\\  In [12] we present a direct algebraic analysis of the
operator gauge invariance condition without using the identities between
C-number distributions.\\
\\
Normalizability of the theory is the second important property. In the causal
approach the question of the normalizability of a quantum field theory does not
require proving its finiteness. Ultraviolet Divergences do not appear at all in
our approach. The problem of normalizability means  that we have to show that
the number of the  finite constants to be fixed by physical conditions stays
the same in all orders  of perturbation theory. This means that finitely many
normalization conditions are sufficient to determine the S-matrix completely.
In the causal approach, the question of  normalizability is totally separated
from the analysis of gauge invariance.\\
The concept of the singular order of distributions (see Appendix B) is a
rigorous definition of the usual power-counting degree [4]. The singular order
$\omega$ depends on the external field operators only so that there are only
finitely many cases with nonnegative $\omega$, i.e. with free normalization
terms. Therefore, the following theorem establishes the normalizability of the
Yang-Mills theory.\\
\\
{\bf Theorem 2.2 (Normalizability)} In the theory defined by (2.1-2.7) the
singular order $\omega$ of a distribution with $b$ external gluons, $g_u$
external ghost operators, $g_{\tilde{u}}$ anti-ghost operators, $d$ derivatives
on these external operators and f quark or anti-quark pairs, is given by the
following simple expression:
$$\omega \le 4-b-g_u-g_{\tilde{u}} -d-3f \eqno (2.19)$$
This expression is obviously independent of the order $n$ of perturbation
theory.\\
\\
The proof is simply based on rigorous power counting arguments and does not
require any analysis of combinatorial or topological properties of Feynman
graphs. The crucial inputs of the proof are the following properties of the
theory:\\
 (a) The specific coupling $T_1$ (2.1) has a mass dimension smaller or equal
than four and\\
 (b) the singular order of the (anti-) commutator distributions in (2.3),(2.4)
and (2.5) are smaller
than zero.\\
\\
The most important and most subtle property of the S-matrix $S(g)$ is its
unitarity in the physical subspace of the Fock space. The subtlety comes from
the well-known fact that (because of the gauge structure) the gauge boson
sector of the Fock space contains more elements than are physically
distinguishable.\\
As is well-known, the realization of the gauge boson field on a positive
definite Hilbert space $F$ is not possible in a manifestly Lorentz covariant
way: The zeroth component of the gauge boson field must be skew-hermitean, in
contrast to the hermitean spatial components:
$$A_0 = - A_0^+ , \quad A_j = A_j^+ \quad j=1,2,3  \eqno (2.20)$$
where `+` denotes the hermitean conjugation with regard to the positive
definite scalar product of the Fock space $$<\cdot \mid \cdot>: \quad F \times
F \longrightarrow \mbox{C}^+ \eqno (2.21)$$
In addition, one introduces a sesquilinear form in $F$ (an indefinite metric)
defined by a metric tensor  $\eta_A^+ = \eta_A^{-1} = \eta_A$
$$<\cdot \mid \eta_A \cdot >: \quad F \times F \longrightarrow \mbox{C} \eqno
(2.22)$$
In the gauge boson sector of $F$, it is given by
$\eta_A = (-1)^{N^{A_0}} $,
where $N_{A_0}$ is the particle number operator of the scalar gauge bosons.
The corresponding conjugation `$k$` for any operator $\hat{O}$ is given by
$$\hat{O}^k = \eta \hat{O}^+ \eta. \eqno (2.23) $$ One finds that the gauge
boson field is pseudo-hermitean
$A_\mu^k = A_\mu $.
 One defines a sesquilinear form in the ghost sector of $F$  with $u^k = u$
and  $\tilde{u}^k = -\tilde{u}.$
The specific coupling is then pseudo-hermitean with regard to the sesquilinear
form: $T_1^k = -T_1 = \tilde{T}_1$ This holds for all n-point distributions
$T_n(x)$ by induction [7]:\\
\\
{\bf Theorem 2.3. (Pseudo-Unitarity)}
$$T_n^k(x) = \tilde{T}_n(x) \quad \forall n   \eqno (2.24a)$$
This implies the following  statement about the formal power series:
$$  S^k(g) = S^{-1}(g) \big{)} \eqno (2.24b)$$
 The $\tilde{T}_n(x)$ are the n-point distributions of the inverse
$S^{-1}$-matrix (see Appendix B).\\
\\
Let $N$ be the particle number operator of the unphysical particles: the scalar
and longitudinal vector bosons, and the ghosts. It is a positive self-adjoint
operator with discrete spectrum $n = 0, 1, 2, 3,$..  . The operator $Q$ (2.9)
manifestly does not change the number of unphysical particles. This means that
$N$ commutes with $Q$. Hence the eigenspaces of the operator $N$ for fixed $n$,
 $Eig(N,n)$, are invariant under $Q$ and $Q$ commutes with the corresponding
projection operators. The nullspace Ker$N$ is the physical subspace $F_\bot$ of
transversal gauge bosons. $F_\bot$ is a subspace of Ker$Q$.
We state the definitions:
$$\mbox{Ker}Q: = \{ \alpha \in F \mid Q \alpha = 0\} \eqno (2.25)$$
$$F_\bot: = \{\alpha \in F \mid Q \alpha = 0 \wedge N \alpha = 0\} \eqno
(2.26)$$
$$F_0: = KerQ \cap  (\oplus_{n>0}Eig(N,n)) \eqno (2.27)$$
We call the corresponding projection operators $P, P_\bot$ and $P_0$. One shows
[7]
$$\mbox{Ker}Q = F_\bot \oplus F_0 \quad \mbox{and} \quad F_0 = QF =
\mbox{Range} Q .\eqno (2.28)$$
According to (2.24), we have pseudo-unitarity. The physical unitarity means the
corresponding perturbative relation for the restriction of the S-matrix to the
physical subspace,
$$P_\bot S(g) P_\bot = S_\bot (g) \eqno (2.29)$$
Its inverse is given by
$$\big{(} P_\bot S(g) P_\bot \big{)}^{-1} = \sum_{n} \frac{1}{n!} \int d^4x_1
\ldots \int d^4x_n \quad \tilde{T}_n^{P_\bot} (x_1, \ldots, x_n) g(x_1)\ldots
g(x_n) \eqno (2.30)$$
where the n-point distributions are equal to the following sum over subsets
of\\ X $= \{x_1, \ldots, x_n\}$
$$\tilde{T}_n^{P_\bot}(X) = \sum_{r=1}^{n} (-)^r \sum_{Pr} P_\bot T_{n_1}(X_1)
P_\bot \ldots P_\bot T_{n_r} (X_r) P_\bot. \eqno (2.31)$$\\
\\
{\bf Theorem 2.4. (Physical Unitarity)}
$$\tilde {T}_n^{P_\bot} = P_\bot T_n^+ P_\bot + div \quad \forall n
\eqno(2.32a)$$
where div denotes terms of divergence form as in the condition of gauge
invariance (2.17).  (2.32a) implies the following statement about  a formal
power series:
$$ \quad (S_\bot)(g)^{-1} =S_\bot^+(g) + div(g) \qquad \mbox{where} \qquad
S_\bot = P_\bot SP_\bot \eqno(2.32b)$$
\\
The straightforward inductive proof can be found in [7, Chapter 7]. Therein
physical unitarity is shown as a direct consequence of the operator
gauge invariance condition (2.17) and of the nilpotency of the operator $Q$.\\
 We stress that  in the causal approach the physical infrared problem
is naturally separated by adiabatic switching of the $S$-matrix by a tempered
testfunction $g$ and also absent before the limit $g \rightarrow 1$ is taken.
So all examinations regarding gauge invariance and unitarity are mathematically
well-defined.\\
\vskip1cm
\newpage
  {\large\bf 3 . The Most General Gauge Invariant Coupling }

\vskip1cm
Starting from the Faddeev-Popov coupling $T_1$ (2.1) and leaving out the
unproblematic coupling to matter fields, we search for the most general gauge
invariant coupling $T_1^g$, ie.
$$  {\bf (A)} \quad d_Q T_1^g=div   \eqno (3.1) $$
\vskip1cm
{\bf Lemma 3.1.}: The most general gauge invariant coupling $T_1^g$, $d_Q
T_1^g=div$, which is also invariant under the special Lorentz group {\bf
$L_+^\uparrow$} {\bf (B)} and  under the structure group $G$ {\bf (C)}, which
has ghost number zero - $G(T_1^g)=0$ - {\bf (D)}  and has maximal mass
dimension 4  {\bf (E)} can be written as
$$T_1^g = \alpha_1 \left[ : u_a \tilde{u}_b: \delta_{ab} + \frac{1}{2} :
A_\mu^a A_b^\mu : \delta_{ab} \right] + \eqno(3.2 a.)$$
$$+ \alpha_2 : F_{\mu\nu}^a F_b^{\mu\nu} : \delta_{ab} + \alpha_3
   \varepsilon_{\mu\nu\kappa\lambda} : F_{\kappa\lambda}^a F_b^{\mu\nu} :
\delta_{ab}+\eqno(3.2 b.c.)$$
$$ + \alpha_4 d_Q L_4 + \partial_\mu \sum_{i=5}^{9} \alpha_i L^\mu_i +
\eqno(3.2 d.e.)$$
$$+ \frac{i}{2} gf_{abc} : A_\mu^a A_\nu^b F_c^{\nu\mu} :- igf_{abc} : A_\mu^a
u_b \d^\mu \tilde{u}_c:+ \eqno(3.2 f.g.)$$
$$+ \beta_1 d_Q K_1+\beta_2 \partial_\mu K_2^\mu+ \eqno(3.2 h.i.)$$
$$+ \beta_3 d_Q K_3 + \beta_4 \partial_\mu K_4^\mu + \beta_5 \partial_\mu
K_5^\mu \eqno(3.2 j.k.l.) $$
where
$$L_4 = i :\tilde{u}_a \partial_\kappa A_b^\kappa : \delta_{ab} ; \quad L_5^\mu
=: u_a \partial^\mu \tilde{u}_b : \delta_{ab};$$
$$L_6^\mu = : \partial^\mu u_a \tilde{u}_b : \delta_{ab}; \quad L_7^\mu =:
A_\nu^a \partial^\nu A_b^\mu : \delta_{ab};$$
$$L_9^\mu = : A_a^\mu \partial_\nu A_b^\nu : \delta_{ab};$$
$$K_1 = gf_{abc} : u_a \tilde{u}_b \tilde{u}_c : ;\quad K_2^\mu = igf_{abc} :
A_a^\mu u_b \tilde{u}_c:;$$ $$ \quad K_3 = gd_{abc}: A_\mu^a A_b^\mu
\tilde{u}_c;\quad K_4^\mu = igd_{abc} : A_\kappa^a A_b^\kappa A_c^\mu:;$$
$$K_5^\mu = igd_{abc} : A_a^\mu u_b \tilde{u}_c:$$
\\
The proof  of Lemma 3.1. is given in Appendix A. We make some {\bf remarks} on
this result:\\
\\
$\bullet$ In Appendix A it is explicitly shown that $d_Q T_{n=1}^g$ has the
following representation as a divergence:
$$ d_Q T_{n=1}^g = \partial_\mu \left( T_{1,g}^\mu+B_{1,g}^\mu \right)
\eqno(3.3)$$
where
$$T_{1,g}^\mu=\alpha_1 :u_a
A_b^\mu:\delta_{ab}+\frac{1}{i}\sum_{i=5}^{9}\alpha_i d_Q L_i^\mu+\eqno(3.3
a.b.) $$ $$+ igf_{abc} \left( : A_\nu^a u_b (\partial^\mu A_c^\nu -
\partial^\nu A_c^\mu):-\frac{1}{2}(u_a u_b \partial^\mu \tilde{u}_c:) \right) +
\eqno(3.3 c.d.e.) $$
$$+ \beta_2 \frac{1}{i} d_Q K_2^\mu + \beta_4 \frac{1}{i} d_Q K_4^\mu + \beta_5
\frac{1}{i} d_Q K_5^\mu  \eqno(3.3 f.g.h.) $$
$$ B_{1,g}^\mu=\gamma_1 igf_{abc} \partial_\nu (:u_a A_b^\mu A_c^\nu:), \quad
\partial_\mu B_{1,g}^\mu=0 \eqno (3.4)$$
\\
$\bullet$ Most free constants in $T_1^g$ (3.1) correspond to pure divergences
(terms proportional to $\alpha_5,\ldots,\alpha_9, \beta_2,\beta_4,\beta_5$,
also $\alpha_3$) or to cocycles with respect to the antiderivation $d_Q$ (terms
proportional to $\alpha_4,\beta_1,\beta_3$).
\\
Besides the Faddeev-Popov-coupling $T_1^g$ (2.1) and the latter terms which are
of course automatically gauge invariant in the sense of {\bf (A)}, we have the
quadratic terms proportional $\alpha_1$ and $\alpha_2$. The $\alpha_1$-term
would generate masses of the gauge bosons and of the ghost fields. Note that
the operator gauge invariance condition fixes the relation between the mass
term of the gauge bosons and the ghosts uniquely. It means that if one
introduces masses into the theory perturbatively, the masses are already fixed
by condition {\bf (A)} .\\
The $\alpha_2$-term is the usual kinetic term of the gauge boson. The quadratic
$\alpha_3$-term is a divergence, but it is ruled out by the condition of the
invariance under the discrete symmetries (see below).\\
In the causal formalism, the information of such quadratic terms are already
contained in the fundamental (anti-)commutation relations and the dynamical
equations of the operators. Therefore  we set all quadratic terms in $T^g_1$ to
zero.\\
\\
$\bullet$ The condition $ {\bf (E)}$  - together  with the corresponding
condition for the fundamental (anti-)commutator relations - is sufficient for
the normalizability of the theory
(see section 2).\\
\\
$\bullet$ In addition, we pose the condition of invariance under the discrete
symmetry transformations, parity $P$, charge conjugation $C$ and time inversion
$T$, on $ T_1^g$: In [7, Chapter 5] we established the following (anti-)unitary
representations $U_i$ of the discrete symmetry transformations in the Fock
space which leave invariant the defining equations of the theory in the causal
formalism :
$$U_PA_\mu^a(x) U_p^{-1} = A_a^\mu (x_p), \eqno (3.5)$$
$$ U_Pu_a(x) U_p^{-1} = u_a (x_p),\quad U_P\tilde{u}_a(x) U_p^{-1} =
\tilde{u}_a (x_p)$$
$$U_TA_\mu^a(x) U_T^{-1} = -U^{ab} A_b^\mu(x_T),\eqno (3.6)$$
$$ U_Tu_a(x) U_T^{-1} = -U_{ab} u^b(x_T),\quad U_T\tilde{u}_a(x) U_T^{-1} =
-U_{ab} \tilde{u}^b(x_T)$$
$$U_cA_a^\mu(x) U_c^{-1} = U_{ab} A_b^\mu(x),\eqno (3.7)$$ $$U_cu_a(x) U_c^{-1}
= U_{ab} u_b(x) \quad U_c\tilde{u}_a(x) U_c^{-1} = U_{ab} \tilde{u}_b(x)$$
\\
$U_{ab}$ is defined by the equation $\lambda_a = U_{ab} \lambda_b = -
\bar{\lambda}$ where $\lambda_a$ are the fundamental representation of the
$SU(N)$- generators.
\\
 The condition of invariance under the discrete symmetry transformations posed
on $T_1^g$,

$$ \quad U_c T_1^g (x) U_c^{-1}=T_1^g, \quad U_p T_1^g (x) U_p = T_1^g (x_p),
\quad U_T T_1^g (x) U_T = T_1^g(x_T) \quad {\bf (F)} \eqno (3.8)$$
leads to
$$\alpha_3 = 0, \quad \beta_3=0, \quad \beta_4=0, \quad \beta_5=0   \eqno
(3.9)$$
\\
For the $\alpha_3$-term one should keep in mind that\\
$\partial_\mu^x = \partial^\mu_{x_P}, \quad \partial_\mu^x = -
\partial^\mu_{x_T}, \qquad \epsilon_{\mu\nu\kappa\lambda} = -
 \epsilon^{\mu\nu\kappa\lambda}$ \\in four(3+1) dimensional space-time; for the
$\beta$-terms note \\ $f'=f, \quad d'=-d, \quad \delta '= \delta$.\\
The invariance under the discete symmetry transformations posed on general
$T_n$-distributions and its compatibility with pseudo-unitarity can be
inductively proven [7, Chapter 5].\\
\\
    $\bullet$ The condition of anti-gauge invariance
$$d_{\bar{Q}} T_1^g = div \qquad {\bf (G)} \eqno(3.10)$$

does not give any further restriction: This condition posed on $T_1^g$ in (3.1)
only leads to $\beta_3=0$. But this already results from condition {\bf (F)}
(see 3.9).\\
\\
$\bullet$ As explicitly shown in Appendix A, all Lorentz invariant {\bf (B)},
G-invariant {\bf (C)} terms with ghost number zero {\bf (D)} and with {\bf
four} normal ordered operators which would be compatible with normalizability
{\bf (E)} are ruled out by the gauge invariance condition {\bf (A)}. But the
well-known quadrilinear terms, the four-gluon- and the four-ghost-vertex are
automatically generated in second order of pertubation theory by our gauge
invariance condition as we will show in the following:\\
\\
We study whether gauge invariance, defined in first order by the equation $d_Q
T_{n=1}^g=[Q,T_{n=1}^g]= div \quad$ {\bf (A)}, can be maintained in second
order, $$d_Q T_{n=2}=[Q,T_{n=2}]=div \eqno (3.11)$$ considering the
tree-contributions only. There exists an effective method to reach this goal
[13]. According to Epstein-Glaser method we construct the causal
commutator $$D_{n=2} (x,y) =\bigl[ T_1^g(x),T_1^g(y) \bigr],$$ whose gauge
invariance is a direct consequence of {\bf (A)}:
$$d_Q D_{n=2}(x,y)=[Q,\,[T_1^g(x),\,T_1^g(y)]]=$$
$$=i \d_\nu^x([T_{1,g}^\nu(x),T_1^g(y)])+i
\d_\nu^y([T_1^g(x),T_{1/g}^\nu(y)]).\eqno(3.12)$$
The first term is a divergence with regard to $x$ and the second with regard to
$y$. In fact, the second term is obtained from the first by interchanging $x$
and $y$ and multiplying it by (-1). The question is whether the same
(divergence form) is true for the commutator $[Q,\,R_2(x,y)]$ obtained by
causal splitting of (3.12). Since this commutator agrees with (3.12) on
$\{(x-y)^2\ge 0, x^0-y^0>0\}$, gauge invariance can only be spoiled by local
terms with support $x=y$. But such terms do arise in the process of
distribution splitting of (3.12), so they can be probably removed. \\  Note
that in the splitting of $[Q,\,D_2]$, we have to split only those numerical
distributions which also  appear in $D_2$ because the  commutation does not
affect the numerical distributions in $D_2$. It only changes the field
operators without disturbing normal ordering. With the same convention of
normalization in the splitting of these numerical distributions, we can
calculate $[Q, R_2]$ directly by splitting $[Q, D_2]$. This
procedure has the advantage that it preserves the divergence structure and
shows immediately where gauge invariance may break down.\\
Considering the commutator $[T_{1,g}^\nu(x),T_1^g(y)]$, the splitting of (3.12)
must be performed as follows: We carry out the derivative $\d_\nu^x$ and then
we uniquely split the causal $D$-distributions in each term according to the
formula
$$D(x-y)=\Dr (x-y)-\Da (x-y).\eqno(3.13)$$
Then we have to examine whether the resulting retarded distribution $R_2$ is
again a divergence, that means, whether the derivative $\d_\nu^x$ can again be
taken out after the splitting. This is not the case because
$$\d_\nu\d^\nu \Dr (x-y)=\delta(x-y)\eqno(3.14)$$
in contrast to $\w D(x-y)=0$. This is the only mechanism to spoil gauge
invariance in the tree-contribution. Note that it is only necessary to analyze
the first term in (3.12) since the splitting of the second commutator in (3.12)
leads to local terms with the same sign as in the first term, so that no
compensation is possible.\\
Now it is an easy job to pick up the possible local terms in the splitting
solution $d_Q R_{n=2}\big|_{tree}$ of $\quad d_Q D_{n=2}\big|_{tree}$.\\
There are the following mechanism to get the $\Box$-operator:\\
(a) There is a second derivative $\partial_\nu^x$ in the fermonic coupling
$T_{1/g}^\nu (x)$ or\\
 (b) there is an operator $\partial_\kappa A^\kappa (y)$ in the specific
coupling $T_1^g (y)$, which can be contracted with an operator $A_\nu (x)$ in
$T_{1/g}^\nu (x)$.
\\
Using the formulae (3.1) and (3.3) and taking into account the additional
constraints by the discrete symmetry conditon (3.9), we get the following list
of local terms with four normal ordered operators generated by the terms in
$T_1^g$ and  $T_{1/g}^\nu$ with three normal ordered operators. In the
brackets, we state the corresponding term in the commutator $\partial_\nu^x
\left[ T_{1/g}^\nu, T_1^g \right]$ which leads to the specialized local term:\\

\begin{eqnarray*}
\hspace{1cm} A_{n_1} &=& - g^2 f_{abc} f_{a'b'c} : \partial_\nu u_b A_a^\mu
\delta (x-y) A_\mu^{a'} A_{b'}^\nu : \qquad \left( \partial_\nu^x \left[
(3.3.c),(3.1.f) \right]_{-} \right) \\
A_{n_2} &=& - \frac{1}{2} g^2 f_{bb'c} f_{cac'} : u_b u_{b'} \delta (x-y)
A_\mu^a \partial^\mu \tilde{u}_{c'}: \qquad \left( \partial_\nu^x \left[
(3.3.c),(3.1.g) \right]_{-} \right) \\
A_{n_3} &=& + \frac{1}{2} g^2 f_{bb'c} f_{cac'} : u_b u_{b'} \delta (x-y)
A_\mu^a \partial^\mu \tilde{u}_{c'} : \qquad  \left( \partial_\nu^x \left[
3.3.e),(3.1.g) \right]_{-} \right)\\
A_{n_4} &=& \beta_2 g^2 f_{aba'} f_{a'b'c'} : \partial_\mu \left[  A_a^\mu u_b
\right] \delta (x-y) u_{b'} \tilde{u}_{c'} : \qquad \left( \partial_\nu^x
\left[ 3.3.d),(3.1.i) \right]_{-} \right) \\
A_{n_5} &=& (\beta_2)^2 g^2 f_{abc} f_{ab'c'} : u_{b'} \tilde{u}_{c'} \delta
(x-y) u_b \partial_\sigma A_c^\sigma : \qquad \left( \partial_\nu^x \left[
3.3.f),(3.1.i) \right]_{-} \right)  \\
A_{n_6} &=& - \beta_2 g^2 f_{abc} f_{a'b'a} : \partial_\kappa \left[
A_{a'}^\kappa u_{b'} \right] \delta (x-y) u_b \tilde{u}_{c} : \qquad  \left(
\partial_\nu^x \left[ 3.3.f),(3.1.g) \right] \right) \\
A_{n_7} &=& 2 \beta_1 g^2 f_{abc} f_{a'b'c} : \partial_\mu \left[ A_a^\mu u_b
\right] \delta (x-y) u_{a'} \tilde{u}_{b'}: \qquad \left( \partial_\nu^x \left[
3.3.c),(3.1.h) \right] \right) \\
A_{n_8} &=& -2 \beta_1 g^2 f_{abc} f_{a'b'c} : \partial^\mu \left[ A_\mu^a u_b
\right] \delta (x-y) u_{a'} \tilde{u}_{b'}: \qquad  \left( \partial_\nu^x
\left[ 3.3.d),(3.1.h) \right] \right) \\
A_{n_9} &=& - \beta_1 g^2 f_{abc} f_{a'cc'} : u_a u_b \delta (x-y)
\tilde{u}_{c'} \partial_\kappa A_{a'}^\kappa : \qquad \left( \partial_\nu^x
\left[ 3.3.e),(3.1.h) \right] \right) \\
A_{n_{10}} &=& 4 \beta_1 \beta_2 g^2 f_{c'bc} f_{c'a'b'} : u_{b'}
u_b \delta (x-y) \tilde{u}_c \partial_\sigma A^\sigma_{a'} : \qquad \left(
\partial_\nu^x \left[ 3.3.f),(3.1.h) \right] \right)
\end{eqnarray*}
$$\eqno(3.15)$$

One easily verifies that  $A_{n_7} + A_{n_8} = 0$, $A_{n_2} + A_{n_3} = 0$ and
$A_{n_4} + A_{n_6} = 0 $.
So we have four local terms left in the commutator:

$$\left[ Q, R_{n=2} \right] = div + 2 (A_{n_1} + A_{n_5} + A_{n_9} +
A_{n_{10}}) \eqno (3.16)$$
\\
The factor 2 represents the fact that there is another anomaly contribution of
the second commutator in (3.12).
Using the Jacobi identity, we arrive at

 $$ A_{n_1} + A_{n_5} + A_{n_9} + A_{n_{10}} = g^2 f_{abc'} f_{a'b'c`} :
\partial_\nu u_a A_b^\mu \delta (x-y) A_\mu^{a'} A_{b'}^\nu : +$$
$$ - (
(\beta_2)^2 +2 \beta_1 - 4 \beta_1 \beta_2) \quad
 g^2 f_{ba'c'} f_{b'cc'} : u_{b'} u_b \delta (x-y) \tilde{u}_{c}
\partial_\kappa A_{a'}^\kappa \eqno(3.17)$$

On the other hand, we have tree terms in $R_{n=2}$ with singular order $\omega
\ge 0$. This means that the general splitting solution $\tilde R_{n=2}$\\

$$ \tilde R_{n=2} = R_{n=2} + N_{0_1} + N_{0_2} + N_{0_3} $$

contains the following three unfixed local normalization terms with free
constants $C_1,C_2,C_3$ \\

\begin{eqnarray*} \hspace{2,5cm} N_{0_1} &=& - iC_1 g^2 f_{abc'} f_{a'b'c'}  :
A_\mu^a A_\nu^b \delta (x-y) A_{a'}^\mu A_{b'}^\nu \\
N_{0_2} &=& - iC_2 g^2 f_{ba'c'} f_{b'cc'}  : u_{b'} u_b \delta (x-y)
\tilde{u}_{c} \tilde{u}_{a'} :  \hspace{3,3cm} \mbox{(3.18)} \\
N_{0_3} &=& - iC_3 g^2 f_{abc} f_{a'b'c}  : A_\mu^a A_{a'}^\mu \delta (x-y) u_b
\tilde{u}_{b'}: \end{eqnarray*}

The question whether there exist a gauge invariant splitting solution $\tilde
R_{n=2} \big|_{tree,4}$ is then equivalent to the solvability of the following
anomaly equation:

$$d_Q( N_{0_1} + N_{0_2} + N_{0_3}) - 2( A_{n_1} + A_{n_5} + A_{n_9} +
A_{n_{10}} ) \stackrel{!}{=} div  \eqno (3.19)$$
\\
Calculating the commutators of the normalization terms, one directly shows that
one can fulfill this equation for all $\beta_1$ and $\beta_2$. We have the
following unique solution of the anomaly equation:

$$C_1 = \frac{1}{2}; \quad C_2 = (\beta_2)^2 + 2 \beta_1  - 4 \beta_1 \beta_2 )
;     \quad C_3 = 0 \eqno (3.20)$$
So we arrive at the following lemma:

\vskip1cm
{\bf Lemma 3.2:} The operator gauge invariance condition, $$d_Q T_{n=2}
\big|_{tree,4}= div \quad,$$ uniquely fixes the normalization of $ \quad
T_{n=2} \big|_{tree,4}$ and naturally introduces a four gluon coupling and a
four ghost coupling in $T_{n=2}$:
\begin{eqnarray*} \hspace{2,5cm} N_{0_1} &=& - \frac{1}{2} g^2 i f_{abc'}
f_{a'b'c'}  : A_\mu^a A_\nu^b \delta (x-y) A_{a'}^\mu A_{b'}^\nu
\hspace{3,0cm} \mbox{(3.21)}\\
N_{0_2} &=& - g^2 \left( (\beta_2)^2 + 2 \beta_1 - \beta_1\beta_2 \right) i
f_{ba'c'} f_{b'cc'}  : u_{b'} u_b \delta (x-y) \tilde{u}_{c} \tilde{u}_{a'} :
.\end{eqnarray*}
\\
\\
{\bf Remark}:\\
  One can directly compare equations (3.1), (3.9), (3.21) and (3.22) with the
most general Lagrangian ( written in terms of interacting field operators )
which is invariant under the full BRS-transformations of the interacting fields
and fulfills reasonable certain additional conditions like {\bf(B)-(G)}. For
example see [14, formula (3.13)]:\\
 Taking into
account the conventions in [14] and setting the parameters $\lambda$  and
$\alpha$  in formula (3.13) of [14] to:\\
 $$\lambda = 1 \mbox{(Feynman gauge)}, \qquad \alpha = \beta_2 =  2 \beta_1
\eqno(3.22)$$\\
 one verifies in a straightforward comparision that in the perturbative
analysis the interaction terms of this Lagrangean, the $g-$ and also
the $g^2-$terms, agree with the  terms in $T^g_1$ and the local normalization
terms in $T^g_2$ (3.21) (Note the $\frac{1}{n!}$ - factor in (1.1)!).\\
Once again one can realize that the linear operator condition of nonabelian
gauge invariance (2.17) is sufficient to derive the whole content of nonabelian
gauge symmetry in perturbation theory.\\
\\
The theorems (2.1) - (2.4) about nonabelian gauge invariance, normalizability,
pseudo-unitarity and physical unitarity hold also in the generalized theory
defined by $T_1^g$. \\
The proof of normalizability can be taken over without any changes,
taking into account that the two crucial inputs are not changed:  the maximal
mass
dimension of the specific coupling and the singular order of the (anti-)
commutation distributions .\\ The algebraic analysis of the nonabelian gauge
invariance condition [12] also covers the generalized theory defined by
$T_1^g$. The condition
of pseudo-hermiticity of $T_1^g$

$$(T_1^g)^k = -T_1^g = \tilde{T}^g_1  \eqno(3.23)$$
poses some additional restrictions on the parameter set:
$$ Re \alpha_i = 0, \quad  Im \beta_i = 0 \eqno(3.24)$$

The proof of the  pseudo-unitarity and of the physical unitarity can then  be
taken over without any changes, as well.\\

\vskip1cm
{\large\bf Section 4 Linear $\xi$- Gauges}
\vskip1cm

The analysis so far has been carried out in the Feynman gauge. In this section,
we discuss the required generalization to other gauge fixings. In this context,
we focus on the so-called linear $\xi$-gauges.
\\
The defining equations of the theory are modified accordingly:\\
\\
$\bullet$  The wave equation of the gauge boson field in the Feynman gauge

$$\Box A_\mu = 0 \quad (\xi = 1)  \eqno (4.1)$$
generalizes to
$$\Box A_\mu - \frac{(\xi -1)}{\xi} \partial_\mu \partial_\kappa A^\kappa = 0
\eqno (4.2)$$
$\bullet$  The commutation relation of the gauge boson field in die Feynman
gauge is
$$\left[ A_\mu (x), A_\nu (y) \right] = ig_{\mu\nu} D_0 (x-y) \qquad  (\xi = 1)
 \eqno (4.3)$$
The Pauli-Jordan distribution $D_0$ is the only one of the well-known two
linear independent Lorentz invariant solutions of the wave equations which is
causal, that means $D_0(x)=0$ for $x^2<0$. The right side of (4.3) thus
represents the general Lorentz invariant and causal ansatz which is compatible
with the manifest normalizability of the theory - the singular order of $D_0$
is smaller than zero. We search for the corresponding general ansatz in the
case of general $\xi$.
\\
Here one should keep in mind that (4.2) implies the dipole equation
$$\Box^2 A_\mu (x) = 0  \eqno (4.4)$$

If $F$ is a causal and Lorentz invariant distribution fulfilling the dipole
equation, we have $\Box F(x) = D_0 (x)$ because supp $\Box F(x) \subseteq$ supp
$F(x)$. We arrive at the following  general ansatz for the commutator
distribution in a linear $\xi$-gauge compatible with causality {\bf (I)},
Lorentz invariance {\bf (II)}, and equation (4.4) {\bf (III)}.

\begin{eqnarray*} \hspace{2,5cm} \left[ A_\mu (x), A_\nu (y) \right]_{-} &=&
ig_{\mu\nu} D(x-y) + i \alpha \partial_\mu \partial_\nu D(x-y)\\
& & \hspace{-0,4cm} \hbox{}+ i \beta g_{\mu\nu} E(x-y) + i \gamma \partial_\mu
\partial_\nu E(x-y) \end{eqnarray*}
$$\eqno(4.5)$$
where $\alpha,\beta,\gamma$ are free constants and $E(x)$ the well-known dipole
distribution

$$E(x) = \int sgn(p_0) \delta' (p^2) e^{-ipx} d^4 p  \eqno (4.6)$$

with the properties

$$\Box^2 E(x) = 0, \quad \Box E(x) = D_0, \quad E(x) = 0 \quad \mbox{for} \quad
x^2 < 0$$

Note that the positive and negative frequency parts of $\hat{E} (p) \quad
\hat{E}^\pm (p) \sim \Theta (\pm p_0) \delta' (p^2)$ are not uniquely defined
because of the indeterminacy of the product $\Theta(\pm p_0) \delta'(p^2)$.
However, the product $sgn(p_0)  \delta'(p_2)$ is a well-defined distribution as
an odd homogeneous tempered distribution (see [15] for details). This point
indicates a well-known difficulty in a general $\xi$-gauge [15,16] which we
will discuss in a forthcoming note.
\\
The additional condition of manifest normalizability {\bf (IV)} leads to
$\alpha = 0$ in (4.5). Furthermore, the required compatibility with equation
(4.2) {\bf (V)} imply $\beta = 0$ and $\gamma = (\xi -1)$. So the second
defining equation of the theory in a general linear $\xi$-gauge reads as
follows

$$\left[ A_\mu (x), A_\nu (y) \right]_{-} = ig_{\mu\nu} D_0 (x-y) + (\xi -1) i
\partial_\mu \partial_\nu E(x-y)  \eqno (4.7)$$
\\

$\bullet$  Generalizing the formula (2.9), we now define the generator of the
linear gauge transformations

$$Q:= \int d^3 x \quad \frac{\partial_\nu A_a^\nu(x)}{\xi}
\stackrel{\leftrightarrow}{\partial}_0 u (x)  \eqno (4.8)$$
Note that we still have the crucial property: $Q^2 = 0$ for all $\xi$.
(4.8) implies for the corresponding antiderivation $d_Q$:

$$d_Q A_\mu^a = i \partial_\mu u_a; \quad d_Q \tilde{u}_a (x) = -i
\frac{\partial_\nu A^\nu}{\xi};$$
$$d_Q \partial_\mu \tilde{u}_a = -i \partial_\mu \partial_\nu A^\nu
\frac{1}{\xi} = i \partial_\nu F^{\nu\mu}; \quad d_Q u_a = 0. \eqno(4.9)$$
Now we further pursue the standard procedure in the causal formalism and
construct the most general gauge invariant specific coupling in a general
$\xi-$gauge.\\
\\
{\bf Lemma 4.1:} The most general gauge invariant coupling $T_1^g$  for a
general linear $\xi$-gauge, {\bf (A)} $\quad d_Q T_1^g = div \quad$  , which
fulfills the conditions {\bf (B)-(G)} agree with the result for $\xi = 1$
(Feynman gauge) (see formulae (3.1/3.9)) with one minor change in the
$\alpha_1$-term:
$$T_1^g = \alpha_1 \left[ \xi : u_a \tilde{u}_b: \delta_{ab} + \frac{1}{2} :
A_\mu^a A_b^\mu : \delta_{ab} \right] + \eqno(4.10)$$
$$\alpha_2 : F_{\mu\nu}^a F_b^{\mu\nu} : \delta_{ab} + \alpha_4 d_Q L_4 +
\partial_\mu \sum_{i=5}^{9} \alpha_i L^\mu_i + $$
$$+ \frac{i}{2} gf_{abc} : A_\mu^a A_\nu^b F_c^{\nu\mu} :- igf_{abc} : A_\mu^a
u_b \d^\mu \tilde{u}_c:+ $$
$$+ \beta_1 d_Q K_1+\beta_2 \partial_\mu K_2^\mu$$

where
$$L_4 = i :\tilde{u}_a \partial_\kappa A_b^\kappa : \delta_{ab} ; \quad L_5^\mu
=: u_a \partial^\mu \tilde{u}_b : \delta_{ab};$$
$$L_6^\mu = : \partial^\mu u_a \tilde{u}_b : \delta_{ab}; \quad L_7^\mu =:
A_\nu^a \partial^\nu A_b^\mu : \delta_{ab};$$
$$L_9^\mu = : A_a^\mu \partial_\nu A_b^\nu : \delta_{ab};$$
$$K_1 = gf_{abc} : u_a \tilde{u}_b \tilde{u}_c : ;\quad K_2^\mu = igf_{abc} :
A_a^\mu u_b \tilde{u}_c:;$$
\\
The proof is straightforward and analogous to the corresponding one in section
3. The explicit representation of $d_Q T_1^g$ as a
divergence reads
$$ d_Q T_{n=1}^g = \partial_\mu \left( T_{1,g}^\mu+B_{1,g}^\mu \right)
\eqno(4.11)$$
where
$$T_{1,g}^\mu=\alpha_1 :u_a
A_b^\mu:\delta_{ab}+\frac{1}{i}\sum_{i=5}^{9}\alpha_i d_Q L_i^\mu+$$
$$igf_{abc} \left( : A_\nu^a u_b (\partial^\mu A_c^\nu -  \partial^\nu
A_c^\mu):-\frac{1}{2}(u_a u_b \partial^\mu \tilde{u}_c:) \right) + \beta_2
\frac{1}{i} d_Q K_2^\mu $$
$$B_{1,g}^\mu=\gamma_1 igf_{abc} \partial_\nu (:u_a A_b^\mu A_c^\nu:), \quad
\partial_\mu B_{1,g}^\mu=0$$

The $\alpha_1$-term again would generate masses of the gauge bosons and of the
ghost fields. Note that the operator gauge invariance condition fixes the
relation between the mass term of the gauge bosons and the ghosts uniquely.
This relation now depends on the parameter $\xi$. Again we leave out the
quadratic terms in the following.\\
Again the gauge invariance condition {\bf (A)} rules out all terms with four
normal ordered operators. But the quadrilinear terms are automatically
generated in second order of perturbation theory by the operator gauge
invariance condition:\\
 The condition $d_Q T_{n=2} \big|_{tree,4} = div$ fixes all free
normalization terms in second order of perturbation theory.\\
In fact, using the procedure of section 3 we can calculate the local terms in
the splitting solution of the commutator  $\partial_\nu^x \left[ T_{1/g}^\nu,
T_1^g \right]$ in a general $\xi$-gauge. Because of the second term in the
commutation relation (4.7) there are additional mechanism to get a local term.
Also note that there
are  additional $\frac{1}{\xi}$ factors in $T_{n=1}^g$ and $T_{n=1,g}^\mu$
(namely in  $ \beta_1 d_Q K_1$ and in $\beta_2 \frac{1}{i} d_Q K_2^\mu$). As a
consequence, we get the
following changes compared with the case $\xi=1$ (see (3.15):\\
$$ A_{n_1}^\xi = A_{n_1},  A_{n_2}^\xi = A_{n_2}, A_{n_3}^\xi = A_{n_3},
A_{n_4}^\xi =  \xi A_{n_4}, A_{n_5}^\xi = A_{n_5},\eqno(4.12)$$
$$ A_{n_6}^\xi = A_{n_6}, A_{n_7}^\xi = (\xi)^{-1} A_{n_7}, A_{n_8}^\xi =
A_{n_8}, A_{n_9}^\xi = (\xi)^{-1} A_{n_9}, A_{n_{10}}^\xi = (\xi)^{-1}
A_{n_{10}}. $$
Moreover, we have an additional local term in $\left( \partial_\nu^x \left[
3.3.c),(3.1.i) \right]_{-} \right)$:
$$ A_{n_{11}}^\xi = - \beta_2 g^2 (\xi - 1) f_{aba'} f_{a'b'c'} : \partial_\mu
\left[  A_a^\mu u_b \right] \delta (x-y) u_{b'} \tilde{u}_{c'} :\eqno(4.13)$$
\\
One easily verifies the cancellations:    $$A_{n_7}^\xi + A_{n_8}^\xi = 0,\quad
 A_{n_2}^\xi + A_{n_3}^\xi = 0$$ $$ A_{n_4}^\xi + A_{n_6}^\xi + A_{n_{11}}^\xi
= 0. \eqno(4.14)$$
We have four local terms left in the commutator again:
 $$ A_{n_1}^\xi + A_{n_5}^\xi + A_{n_9}^\xi + A_{n_{10}}^\xi = g^2 f_{abc'}
f_{a'b'c`} : \partial_\nu u_a A_b^\mu \delta (x-y) A_\mu^{a'} A_{b'}^\nu : +$$
$$ - (
(\beta_2)^2 +2 \beta_1 (\xi)^{-1} - 4 \beta_1 \beta_2  (\xi)^{-1} ) \quad
 g^2 f_{ba'c'} f_{b'cc'} : u_{b'} u_b \delta (x-y) \tilde{u}_{c}
\partial_\kappa A_{a'}^\kappa \eqno(4.15)$$
As in the $\xi =1$ case we have three unfixed local normalization terms
$N_{0_1}, N_{0_2}, N_{0_3}$ with free constants $C_1,C_2,C_3$ (see (3.18)). The
question whether there exist a gauge invariant 2-point distribution $\tilde
T_{n=2} \big|_{tree,4}$ is equivalent to the solvability of the following
anomaly equation:

$$d_Q( N_{0_1} + N_{0_2} + N_{0_3}) - 2( A_{n_1} + A_{n_5} + A_{n_9} +
A_{n_{10}} ) \stackrel{!}{=} div  \eqno (4.16)$$
\\
For any $\beta_1$ and $\beta_2$ we can find the following  (unique) solution:

$$C_1 = \frac{1}{2}; \quad C_2 = ((\beta_2)^2 +2 \beta_1 (\xi)^{-1} - 4 \beta_1
\beta_2  (\xi)^{-1} );     \quad C_3 = 0 \eqno (4.17)$$
So we have also in a general $\xi$-gauge:
\vskip1cm
{\bf Lemma 4.2:} The operator gauge invariance condition, $$d_Q T_{n=2}
\big|_{tree,4}=div\quad,$$ uniquely fixes the normalization of $ \quad T_{n=2}
\big|_{tree,4}$ and naturally introduces a four gluon coupling and a four ghost
coupling in $T_{n=2}$:
$$ N_{0_1} = - \frac{1}{2} g^2 i f_{abc'} f_{a'b'c'}  : A_\mu^a A_\nu^b \delta
(x-y) A_{a'}^\mu A_{b'}^\nu  : \eqno(4.18)$$
$$ N_{0_2} = - g^2   ((\beta_2)^2 +2 \beta_1 (\xi)^{-1} - 4 \beta_1 \beta_2
(\xi)^{-1} )   i f_{ba'c'} f_{b'cc'}  : u_{b'} u_b \delta (x-y) \tilde{u}_{c}
\tilde{u}_{a'} :$$
\\
The comparison with the Lagrange formalism [14] leads to the same conclusion as
in section 3: We consider the most general Lagrangian in a general $\xi$-gauge
( written in terms of interacting field operators ) which is invariant under
the full BRS-transformations of the interacting fields and fulfills reasonable
certain additional conditions like {\bf(B)-(G)}; for example see again formula
(3.13) in [14]. In perturbative analysis the interaction terms of the
Lagrangian, the $g-$ and also
the $g^2-$terms, agree with the  terms in $T^g_1$ (4.10) and the local
normalization terms in $T^g_2$ (4.18). For an explict comparision one has to
set the parameters  $\lambda$  and $\alpha$  in formula (3.13) of [14] to:
$$\lambda = \xi  \quad \alpha = \beta_2 =  2 \beta_1 (\xi)^{-1} \eqno(4.19)$$\\
\\
The theorems about gauge invariance, normalizability, pseudo-unitarity and
physical unitarity hold also in the theory with a general $\xi$-gauge.  Again,
the proofs can be taken over with minor changes from the $\xi =1$ case. \\
The $\xi$-independence of the $S$-matrix elements is not yet discussed here.
Finally, a short remark on other gauge fixings is in order.\\ The
generalization to nonlinear gauge fixings does not represent a principal
difficulty since one can introduce the nonlinear terms in the specific
coupling. As far as the Coulomb gauge is concerned,
Strocchi and Wightmann have shown that a theory of the free electromagnetic
field that maintains the Maxwell equations as operator identities must use a
vector potential being nonlocal and Lorentz-variant [17].
This is precisely what happens in a gauge theory fixed in the Coulomb gauge.
The specific coupling in the Coulomb gauge contains a nonlocal term. Moreover,
the fundamental commutator also includes a non-causal part. (This missing
microcausality is of course unproblematic because the commutator of the field
strength includes only causal distributions.) It is obvious, that in a
formalism where causality appears as the decisive inductive construction
element, it is difficult to analyze theories in which this very property is not
manifest in their fundamental equations.
\\

\vskip1cm
{\large\bf Conclusions}\\
\\
We have derived the main features of nonabelian gauge theories in four (3+1)
dimensional space time using the causal Epstein-Glaser method. In this
formalism, the technical details concerning the well-known
UV- and IR-problem in quantum field theory are separated and reduced to
mathematically well-defined problems, namely the causal splitting  and the
adiabatic switching of operator-valued distributions.\\
We have shown that the whole analysis of nonabelian gauge symmetry can be done
in the well-defined Fock space of
free asymptotic fields; the LSZ-formalism is not used in our construction.
Nonabelian gauge symmetry is introduced by an operator condition in every order
of perturbation theory separately. The approach allows for a simplified
analysis of the different properties of nonabelian gauge theories .\\

\vskip1,0cm

{\bf Acknowledgement:} I thank G. Scharf for useful discussions, M. Misiak for
careful reading of the manuscript and the Swiss National
Science Foundation for financial support.\\

\vskip1cm
{\large\bf Appendix A}
\vskip1cm

We give a sketch proof of Lemma 3.1:
\\
Because of condition {\bf (E)}, according to which the maximal mass dimension
is four, only terms with two, three or four operators in $T_1^g$ are possible.
\\
$\bullet$ The most general Lorentz invariant, $SU(N)$ invariant local
normalordered local operator-valued distribution  with ghost number zero and
maximal mass dimension four with two basic operators can be written $(\delta_i
\in {\bf C}$ free)

\begin{eqnarray*} \lefteqn{T_1 \big|_2 = \delta_1 : u_a \tilde{u}_b :
\delta_{ab} + \delta_2 : A_\mu^a A_b^\mu : \delta_{ab} \hspace{4,3cm}
\mbox{(A.1)} } \\
& & \mbox{}+ \delta_3 : \partial_\mu u_a \partial^u \tilde{u}_b : \delta_{ab} +
\delta_4 : \partial_\mu A_a^\mu \partial_\kappa A_b^\kappa : \delta_{ab} \\
& & \mbox{} + \delta_5 : \partial_\mu A_\nu^a \partial^\nu A_b^\mu :
\delta_{ab} \mbox{}+ \delta_6 : \partial_\mu A_\nu^a \partial^\mu A_b^\nu :
\delta_{ab}\\
& & \mbox{} + \delta_7 : A_\mu^a \partial_\nu F_b^{\mu\nu} : \delta_{ab} +
\delta_8 \varepsilon_{\mu\nu\kappa\lambda} :F^{\mu\nu}_a F^{\kappa\lambda}_b:
\delta_{ab} \end{eqnarray*}

The most general $SU(N)$ invariant normalordered local operator-valued
distribution with two fundamental operators, ghost number $(-1)$ and maximal
mass dimension four, which is covariant under the adjoint representation of the
Lorentz group, is the following: $(\gamma_i \in {\bf C}$ free)

\begin{eqnarray*} \lefteqn{T_1^\nu \big|_2 = \gamma_1 : u_a A_b^\nu :
\delta_{ab} + \gamma_2 : \partial_\nu u_a \partial_\kappa A_b^\kappa :
\delta_{ab} \hspace{3,4cm} \mbox{(A.2)} } \\
& & \mbox{} + \gamma_3 : \partial_\kappa u_a \partial^\nu A_b^\kappa :
\delta_{ab} + \gamma_4 : \partial_\kappa u_a \partial^\kappa A_b^\nu :
\delta_{ab} \\
& & \mbox{} + \gamma_5 : u_a \partial_\kappa F_b^{\nu\kappa} : \delta_{ab}
\end{eqnarray*}

We search for restrictions on the free parameters $\delta_i \in {\bf C}$ in
(A.1.) by the equation

$$d_Q T_1 (\delta_i) \bigl|_2 = i \partial_\nu T_1^\nu (\gamma_i) \bigr|_2
.\eqno (A.3)$$

(A.3) leads to the following relations between the two parameter-sets
$\gamma_i$ and $\delta_i$:

$$\delta_1 = 2 \delta_2 = \gamma_1 ; \quad 2 ( \delta_5 + \delta_6 ) =
\gamma_3+\gamma_5; \quad \delta_3 + \delta_7 = \gamma_2 + \gamma_4 + \gamma_5
\eqno (A.4)$$

Thus, we have only one restriction on the set $(\delta_i)$, namely $\delta_1 =
2 \delta_2$.
It means, that if one \hyphenation{in-tro-du-ces}introduces masses into the
 theory perturbatively, then gauge invariance already fixes the relation
between the mass of the ghost and the gauge boson at first order.\\
Using (A.4) and making a simple reparametrization one easily arrives at the
two-operator terms in (3.1) and (3.3).\\
\\
$\bullet$ The corresponding ansatz for three basic operators are \\($\quad
\alpha_i,\beta_i,\varepsilon_i,\mu_i \in {\bf C} \quad \mbox{free}$ note that
we have already fixed the coefficient of the three-gluon coupling;
 $\quad f_{abc}$ and $ d_{abc}$ are the well-known totally antisymmetric and
symmetric invariant tensors. These are the only two independent invariant
tensors of rank three for $SU(N), N>2$: for $SU(2)$ we have to set $d=0$.):

\begin{eqnarray*} \lefteqn{T_1 \big|_3 = igf_{abc} \left\{ : A_\kappa^a
A_\lambda^b \partial^\lambda A_c^\kappa : + \varepsilon_1 : A_\kappa^a u_b
\partial^\kappa \tilde{u}_c : + \varepsilon_2 : \partial_\lambda A_a^\lambda
u_b \tilde{u}_c + \varepsilon_3 : A_a^\lambda \partial_\lambda u_b \tilde{u}_c
: \right\} } \\
& & \mbox{} + ig d_{abc} \left\{ \frac{\beta_1}{2} : A_\mu^a A_b^\nu
(\partial_\nu A_c^\mu - \partial^\mu A_\nu^c): + \beta_2 : A_\mu^a A_b^\mu
\partial_\kappa A_c^\kappa : + \beta_3 : A_\mu^a u_b \partial^\mu \tilde{u}_c :
\right. \\
& & \hspace{6,1cm} \biggl. \mbox{}+ \beta_4 : A_\mu^a \partial^\mu u_b
\tilde{u}_c : + \beta_5 : \partial_\mu A_a^\mu u_b \tilde{u}_c:
\biggr\}\end{eqnarray*}

\begin{eqnarray*} \lefteqn{ \hspace{1,5cm} T_1^\nu \big|_3 = igf_{abc} \left\{
\mu_1 : A_\mu^a u_b \partial^\nu A_c^\mu + \mu_2 : A_\mu^a u_b \partial^\mu
A_c^\nu : \right. } \\
& & \hspace{2,7cm} \mbox{} + \mu_3 : u_a u_b \partial^\nu \tilde{u}_c + \mu_4 :
\partial_\nu u_a u_b \tilde{u}_c : \\
& & \hspace{2,7cm} \mbox{} + \left. \mu_5 : A_a^\nu u_b \partial_\kappa
A_c^\kappa : + \mu_6 : \partial^\mu u_a A_\mu^b A_c^\nu : \right\} \\
& & \hspace{1,55cm} \mbox{}+ igd_{abc} \left\{ \alpha_1 : u_a A ^\nu_b
\partial_\mu A_c^\mu: + \alpha_2 : u_a A_\mu^b \partial^\nu A_c^\mu : \right.
\\
& & \hspace{2,7cm} \mbox{} + \alpha_3 : u_a A_\mu^b \partial^\mu A_c^\nu : +
\alpha_4 : \partial_\mu u_a A_b^\nu A_c^\mu : \\
& & \hspace{2,7cm} \mbox{} + \left. \alpha_5 : \partial^\nu u_a A_\mu^b A_c^\mu
: + \alpha_6 : \partial_\nu u_a u_b \tilde{u}_c : \right\} \end{eqnarray*}

$$\eqno(A.5)$$

The condition of gauge invariance:

$$d_Q T_1(\epsilon_i,\beta_i) \big|_3 = i \partial_\nu T_1^\nu(\mu_i,\alpha_i)
\big|_3 \eqno (A.6)$$

leads to the following restrictions on the parameter sets:

$$\alpha_1 = \beta_3, \quad \alpha_1 = \beta_5, \quad \alpha_2 = 0, \quad
\alpha_3 = 0, \quad \alpha_4 = \beta_1, \quad \alpha_5 = \frac{\beta_1}{2}
\eqno (A.7)$$
$$\alpha_6 = \beta_3, \quad \alpha_1 + \alpha_4 = 2 \beta_2 + \beta_4$$
$$\mu_1=1,\quad \mu_2=\mu_6 -1, \quad \mu_3=- \frac{1}{2}, \quad$$
$$ \mu_4=1+\epsilon_1, \quad \mu_5=\epsilon_1 - \mu_2, \quad
\epsilon_3=\epsilon_1 +1 $$

So we have the following restrictions on $T_1 \big|_3$ (A.5):

$$\beta_2 = \frac{1}{2} ( \beta_3 + \beta_1 - \beta_4 ), \quad \beta_5 =
\beta_3, \quad \beta_1,\beta_3,\beta_4 \quad \mbox{still free} \eqno (A.8)$$
$$ \epsilon_3=\epsilon_1 +1, \quad \epsilon_1,\epsilon_2 \quad  \mbox{still
free} $$

With the help of a simple reparametrization we arrive at the representation of
the three-operator terms in (3.1) and (3.3).\\
\\
$\bullet$ The corresponding ansatz for four normalordered operators is \\
 (For $SU(N), N>3$ we have 9 invariant tensors\\ $\delta_{ab}\delta_{cd},
\delta_{ad}\delta_{bc}, \delta_{ac}\delta_{bd},d_{abe} d_{cde}, d_{ade}
d_{cbe}, d_{ace} d_{dbe}, d_{abe} f_{cde}, d_{ade} f_{bce}, d_{ace} f_{dbe}$.
Note that the following ansatz  is also valid for $SU(3)$ , where we have only
eight instead of nine
invariant tensors of rank four due to the well-known relation
$\delta_{ij}\delta_{kl}+\delta_{ik}\delta_{jl}+\delta_{il}\delta_{jk}=3(d_{ijm}
d_{kem}+d_{ikm}d_{jlm}+d_{ilm}d_{jkm})$
. The case $SU(2)$ is straightforward: One has to set $d=0$.):

\begin{eqnarray*} \lefteqn{T_1^\nu \big|_4 = \left( \zeta_1 d_{ace} d_{bde} +
\vartheta_1 \delta_{ac} \delta_{bd} \right) g: u_a u_b \tilde{u}_c \tilde{u}_d
: + } \\
& & \mbox{} + \left( \zeta_2  d_{abe} d_{cde} + \vartheta_2 \delta_{ab}
\delta_{cd} \right) g: A_\mu^a A_b^\mu A_c^\nu A_\nu^d : + \\
& & \mbox{} + \left( \zeta_3  d_{ace} d_{bde} + \vartheta_3 \delta_{ac}
\delta_{bd} \right) g: A_\mu^a A_b^\mu A_c^\nu A_\nu^d : + \\
& & \mbox{} + \left( \zeta_4  d_{abe} d_{cde} + \vartheta_4 \delta_{ab}
\delta_{cd} \right) g: u_a \tilde{u}_b A_\nu^c A_d^nu : + \\
& & \mbox{} + \left( \zeta_5  d_{ace} d_{bde} + \vartheta_5 \delta_{ac}
\delta_{bd} \right) g: u_a \tilde{u}_b A_\nu^c A_d^\nu : + \\
& & \mbox{} + \left( \kappa_1 d_{abe} f_{cde} + \kappa_2 d_{ace} f_{bde}
\right) g : A_\mu^a A_b^\mu u_c \tilde{u}_d : + \\
& & \mbox{} + \left( \kappa_3 d_{ade} f_{bce} \right) g : A_\mu^a A_b^\mu u_c
\tilde{u}_d : +\\
& & \mbox{} + \left( \kappa_4 d_{ace} f_{bde} \right) g : u_a u_b \tilde{u}_c
\tilde{u}_d :
\end{eqnarray*}

\begin{eqnarray*} \lefteqn{T_1^\nu \big|_4 = \left( \lambda_1 d_{ace} d_{bde} +
\mu_1 \delta_{ac} \delta_{bd} \right) g: u_a u_b \tilde{u}_c A_d^\nu : + } \\
& & \mbox{} + \left( \lambda_2  d_{abe} d_{cde} + \mu_2 \delta_{ab} \delta_{cd}
\right) g: u_a A_b^\nu A_c^\mu A_\mu^d : + \\
& & \mbox{} + \left( \lambda_3  d_{ace} d_{bde} + \mu_3 \delta_{ac} \delta_{bd}
\right) g: u_a A_b^\nu A_c^\mu A_\mu^d : + \\
& & \mbox{} + \left( \nu_1  d_{ace} f_{bde} + \nu_2 f_{ace} d_{bde} \right) g:
u_a u_b \tilde{u}_c A_d^\nu : + \\
& & \mbox{} + \left( \nu_3  f_{abe} d_{cde} \right) g: u_a u_b \tilde{u}_c
A_d^\nu : + \\
& & \mbox{} + \left( \nu_4  f_{abe} d_{cde} \right) g: u_a A_b^\nu A_\mu^c
A_d^\mu : + \\
& & \mbox{} + \left( \nu_5  f_{ace} d_{bde} + \nu_6 d_{ace} f_{bde} \right) g:
u_a A_b^\nu A_\mu^c A_d^\mu : \end{eqnarray*}
$$\eqno(A.9)$$
$(\zeta_i,\vartheta_i,\kappa_i,\lambda_i,\mu_i,\nu_i \quad \in {\bf C}$ free
constants).\\
 The equation $$d_Q T_1 \left( \zeta_i,\vartheta_i,\kappa_i \right) \big|_4 = i
\partial_\nu T_1^\nu \left( \lambda_i, \mu_i, \nu_i \right) \big|_4
\eqno(A.10)$$ has no nontrivial solution.  As a consequence, there are no
quadrilinear terms in the specific coupling $T^g_1$. \hfill \hbox{}\\

\vskip1cm
{\large\bf Appendix B}
\vskip1cm
In this appendix we state some main features of the Epstein-Glaser method in
quantum field theory (for details see [1,4]):\\
Epstein and Glaser followed the Bogoliubov's formalism in order to keep apart
the different difficulties encountered in perturbative quantum field theory. In
contrast to the usual Lagrangean approach, Epstein and Glaser construct the
perturbative scattering matrix $S(g)$ directly in the well-defined Fock space
of free fields $F$. In order to obtain the explicit form of the $S$-matrix,
they use certain physical conditions. Besides Poincare invariance the condition
of causality plays the most important role:\\

$\bullet$ If the support of $g_{1}\epsilon\cal S$ in Minkowski space is earlier
than the support of $g_{2}\epsilon\cal S$ in some Lorentz frame (supp$g_{1}<$
supp$g_{2}$), then the S-matrix fulfills the following functional equation:
$$S(g_1+g_2)=S(g_2)\cdot S(g_1)\quad \mbox{\bf[Causality (I)]} \eqno(B.1)$$
Note that we use a slightly different condition of causality compared with
the one used by Epstein and Glaser.\\
$\bullet \quad U(a, \Lambda)$ shall be the usual representation of the
Poincar\'e group $P_+^\uparrow$ in the Fock space $F$. The condition of {\bf
Poincar\'e invariance} of the S-matrix can be expressed as follows:
$$U(a,\bf 1\it )S(g)U(a,\bf 1\it )^{-1} = S(g_a) \quad \forall a \epsilon \bf
R^{\it 4} $$
$$\mbox{where} \quad g_a(x)=g(x-a).\quad \hbox{\bf[Translational invariance
(II)]} \eqno(B.2)$$
$$U(0,\Lambda)S(g)U(0,\Lambda)^{-1}=S(g_\Lambda), \quad \forall \Lambda \in
L_+^\uparrow $$
	$$\mbox{where} \quad g_\Lambda(x)=g(\Lambda^{-1}x).\quad \hbox{\bf[Lorentz
Invariance (III)]} \eqno(B.3)$$
$\bullet$ Epstein and Glaser search for the most general Poincare invarinat
solution of the functional equation for the S-matrix of the following form
(formal power series in $g\epsilon\cal S$)
$$S(g)=\bf 1 \it + \sum_{n=1}^\infty \frac{1}{n!} \int d^4x_1 \ldots d^4 x_n
T_n(x_1,\ldots x_n)g(x_1)\ldots g(x_n)$$
$$ \=d \bf 1 \it + \it T.\rm \quad \hbox{\bf[Perturbative Ansatz (IV)]
}\eqno(B.4)$$
if  the {\bf Specific Coupling} of the theory $T_{n=1}$ (\bf V \rm) is given.
The $T_{n}$ are operator-valued n-point distributions.\\
Epstein and Glaser show that the whole perturbative S-matrix in the sense of a
formal power series (IV) is already determined by the conditions of causality
(I), translational invariance (II) and the specific coupling of the theory (V)
except for a number of finite (!) free constants which have to be fixed by
further physical conditions. The main steps of their inductive construction are
the following:\\
$\bullet$ Analogously to (B.4), Epstein and Glaser express the inverse S-matrix
also by a formal power series:
$$S(g)^{-1}=1+\sum_{n=1}^\infty \frac {1}{n!} \int d^4x_{1}\ldots d^4 x_n
\tilde{T}_n(x_1,\ldots x_n)g(x_1)\ldots g(x_n)$$
$$ =(\bf 1 \rm + \it T)^{-1}=\bf 1\rm + \sum_{n=1}^\infty (-\it
T)^r.\eqno(B.5)$$
Since by definition $\tilde{T}(x_1, \ldots , x_n)$ and also $T_n (x_1, \ldots ,
x_n)$ are symmetric in $x_1, \ldots , x_n$, it is convenient to use a
set-theoretical notation X = {$x_1, \ldots , x_n$}. The distributions
$\tilde{T}$ can be computed by formal inversion of (B.4):
$$\tilde{T}_n(X)=\sum_{r=1}^n(-)^r\sum_{P_r}T_{n_1}(X_1)\ldots
T_{n_r}(X_r),\eqno(B.6)$$
where the second sum runs over all partitions $P_r$ of X into r disjoint
subsets
$$X=X_1\cup\ldots\cup X_r,\quad X_j\not=\emptyset,\quad \mid X_j \mid
=n_j.\eqno(B.7)$$
We stress the fact that all products of distributions are well-defined because
the arguments are disjoint sets of points so that the products are tensor
products of distributions.

$\bullet$ Epstein and Glaser translate the conditions imposed on $S(g)$ into
the corresponding perturbative conditions on the n-point distributions $T_n
(x_1,\ldots ,x_n)$ and $\tilde{T_n}(x_1,\ldots ,x_n) \quad n\epsilon \bf N
\it$, according to the Bogoliubov's approach.

$\bullet$ Now Epstein and Glaser introduce the retarded and the advanced
n-point distributions:
$$R_n(x_1,\ldots ,x_n)=T_n(x_1,\ldots ,x_n)+R'_n \qquad \mbox {where}\quad
R'_n=\sum_{P_2}T_{n-n_1}(Y,x_n)\tilde{T}_{n_1}(X) \eqno(B.8)$$
$$A_n(x_1,\ldots ,x_n)=T_n(x_1,\ldots ,x_n)+A'_n \qquad \mbox{where} \quad
A'_n=\sum_{P_2}\tilde{T}_{n_1}(X)T_{n-n_1}(Y,x_n). \eqno(B.9)$$
The sum runs over all partitions $P_2:\{x_1,\ldots x_{n-1} \}=X \cup Y, \quad X
\not= \emptyset$ into disjoint subsets with $\mid X \mid =n_1 \ge 1, \mid Y
\mid \le n-2.$

Both sums, $R'_n$ and $A'_n$, contain $T_j$'s with $j \le n-1$ only and are
therefore known quantities in the inductive step from $n-1$ to $n \quad$ - in
contrast to $T_n$.

Note that the last argument $x_n$ is marked as the reference point for the
support of $R_n$ and $A_n$.\\ The following proposition is a consequence of the
causality condition {\bf (I)}:
\\
{\bf Proposition B.1}
$$\mbox{supp} R_m(x_1,\ldots ,x_m) \subseteq \Gamma_{m-1}^+(x_m),
\quad m < n $$
$$\mbox{supp} A_m(x_1,\ldots ,x_m) \subseteq \Gamma_{m-1}^-(x_m),
\quad m < n \eqno(B.10)$$
where $\Gamma_{m-1}^+$ ($\Gamma_{m-1}^-$) is in the (m-1)-dimensional closed
forward (backward) cone
$$\Gamma_{m-1}^+(x_m)=\{(x_1,\ldots ,x_{m-1}) \mid (x_j - x_m)^2 \ge 0, x_j^0
\ge x_n^0, \forall j \}.$$
\\
In the difference
$$D_n(x_1, \ldots ,x_n) \=d R_n-A_n = R'_n-A'_n\eqno(B.11)$$
the unknown n-point distribution $T_n$ cancels. Hence this quantity is also
known in the inductive step. It should be added that $D_n$ has a causal
support:\\
{\bf Proposition B.2}
$$\mbox{supp} D_n \subseteq \Gamma_{n-1}^+(x_n) \cup \Gamma_{n-1}^-(x_n)
\eqno(B.12)$$
This crucial support property is preserved in the inductive step. It directly
results from causality.

$\bullet$ Given the aforegoing facts, the following inductive construction of
the n-point distribution $T_n$ becomes possible:

Starting off with the known $T_m(x_1, \ldots , x_n), m \le n-1,$ one computes
$A'_n, R'_n$ and $D_n = R'_n - A'_n.$ With regard to the supports, one can
decompose $D_n$ in the following way:
$$D_n(x_1, \ldots , x_n) = R_n (x_1, \ldots , x_n) - A_n (x_1, \ldots ,
x_n)\eqno(B.13)$$
$$\mbox{supp} R_n \subseteq \Gamma_{n-1}^+(x_n), \quad \mbox{supp} A_n
\subseteq \Gamma_{n-1}^-(x_n)$$
Then $T'_n$ is given by
$$T'_n = R_n - R'_n = A_n - A'_n \eqno(B.14)$$
One can verify that the $T'_n$ satisfy the perturbative versions of conditions
[1].\\
Because of the marked $x_n$-variable, we finally symmetrize:
$$T_n(x_1,\ldots x_n)=\sum_{\pi} \frac{1}{n!} T'_n(x_{\pi 1}, \ldots x_{\pi n})
\eqno(B.15)$$
The only nontrivial step in the construction is the splitting of the
operator-valued distribution $D_n$ with support in $\Gamma^+ \cup \Gamma^-$
into a distribution $R_n$ with support in $\Gamma^+$ and a distribution $A_n$
with support in $\Gamma^-$. In causal perturbation theory the usual
renormalization program is reduced to this conceptually simple and
mathematically well-defined problem:\\
$\bullet$  Let there be an operator-valued tempered
\hyphenation{dis-tri-bu-tion} distribution with causal support:
$$D\epsilon \cal S' (\bf R^{\it 4n}),\quad \it supp D \subset \Gamma^+ (x_n)
\cup \Gamma^- (x_n) \eqno (B.16)$$
The question is whether it is possible to find a pair (R, A) of tempered
distributions on $\bf R^{\it 4n}$ with the following characteristics:
$$\bullet\quad R, A \epsilon \cal S \it' (\bf R^{\it 4n})\qquad \mbox{\bf
(A)}$$
$$\bullet\quad R \subset \Gamma^+(x_n), \quad  A \subset \Gamma^-(x_n)\qquad
\mbox{\bf (B)} \eqno (B.17)$$
$$\bullet\quad R - A = D\qquad \mbox{\bf (C)}$$
Because of $\Gamma^+_{n-1} (x_n) \cap \Gamma^-_{n-1} (x_n) =
\{(x_n,...,x_n)\}$, it is obvious that the behaviour of the distribution at $x
= (x_n,....,x_n)$ is crucial for the splitting problem. One has to classify the
singularities of distributions in this region. This can be a carried out with
the help of the singular order of distributions which is a rigorous definition
of the usual power-counting degree. We finally state the main definitions:\\
We assume $d(x)$ to be a tempered distribution in $ \cal S'\it (\bf R^{\it
m}),{\it m} = 4({\it n}-1)$.\\
{\bf Definition B.1}  The distribution $d(x) \epsilon \cal S' \it(\bf R^{\it
m})$ has quasi-asymptotics $d_0(x)$ at $x = 0$, with regard to a positive
continuous function $\rho(\delta), \delta > 0$ if the limit
$$\lim_{\delta\rightarrow 0} \rho(\delta)\delta^m d(\delta x) = d_0(x) \neq 0
\eqno (B.18)$$
exists in $S'(R^m)$.\\
By scaling transformation it follows that
$$\lim_{\delta\rightarrow 0} \frac{\rho(a\delta)}{\rho(\delta)}=a^\omega \eqno
(B.19)$$
with some real $\omega.$ $\rho$ is called power-counting function.\\
{\bf Definition B.2}  The distribution $d(x))\in \cal S' \rm (\bf R^{\it m})$
is called singular of order $\omega$ at $x = 0)$, if it has a quasi-asymptotics
$d_0(x))\quad at \quad x = 0$ with power-counting function $\rho(\delta)$
satisfying (B.19).

Note that this definition differs from the one introduced by Epstein and Glaser
 [1]. The latter definition is hampered by the fact that the corresponding
definitions in the $x$-space and $p$-space are not completely equivalent. Our
definition does not have this defect.

\vspace{2cm}

{\Large\bf References}

\vspace{0.5cm}

\begin{tabbing}

1. \quad\quad\quad\=  H. Epstein, V. Glaser,\\
 \>  in G. Velo, A.S. Wightman (eds.):\\
 \>  Renormalization Theory,\\
 \>  D. Reidel Publishing Company, Dordrecht 1976, 193\\
  \> H. Epstein, V. Glaser, \\
 \> Annales de l'Institut Poincare 29 (1973) 211\\
2. \> M. D\"utsch, F. Krahe, G. Scharf, \\
 \> Nuovo Cimento 103A (1990) 903\\
3. \> R. Seneor,\\
 \>  in G. Velo, A.S. Wightman (eds.):\\
 \>  Renormalization Theory,\\
 \>  D. Reidel Publishing Company, Dordrecht 1976, 255\\
4. \> G. Scharf,\\
 \>  Finite Quantum Electrodynamics (Second Edition),\\
 \>  Springer, Berlin 1995\\
5. \> M. D\"utsch, T. Hurth, G. Scharf,\\
 \> Physics Letters B327 (1994) 166\\
6. \> M. D\"utsch, T. Hurth, G. Scharf, \\
 \>  Nuovo Cimento 108A (1995) 679, 108A (1995) 737 \\
7. \> T. Hurth,\\
 \>  Annals of Physics 244 (1995) 340, hep-th/9411080\\
8. \>  C. Becchi, A. Rouet, R. Stora,\\
 \>  Annals of Physics 98 (1976) 287\\
9. \> R. Delbourgo, S. Twisk, G. Thompson,\\
\> International Journal of Modern Physics A3 (1988) 435\\
\> A. Burnel,\\
\> Physical Review D33 (1986) 2981, D33 (1986) 2985\\
10. \> T. Hurth,\\
 \>  Z\"urich University Preprint ZU-TH-25/95\\
11. \> T. Kugo, I. Ojima, \\
 \> Progress of Theoretical Physics Supplement 66 (1979) 1\\
12. \> T. Hurth,\\
 \>  manuscript in preparation\\
13. \> M. D\"utsch, T. Hurth, F. Krahe, G. Scharf,\\
 \>  Nuovo Cimento 106A (1993) 1029\\
14.  \> L. Baulieu,\\
 \>  Physics Reports 129 (1985) 1\\
15. \> N.N. Bogoliubov, A.A. Logunov, A.I. Oksak, I.T. Todorov,\\
 \>  General Principles of Quantum Field Theory,\\
 \> Kluwer Academic Publishers, Dordrecht 1990\\
16.  \> P.J.M. Bongaarts,\\
 \> Journal of Mathematical Physics 23 (1982) 1881\\
17. \> F. Strocchi, A.S. Wightman,\\
\> Journal of Mathematical Physics 15 (1974) 2198 \\

\\
\\
\end{tabbing}

\end{document}